\documentclass[aps, twocolumn, superscriptaddress]{revtex4}

\makeatletter
\def\@dotsep{4.5}
\makeatother

\usepackage{graphicx}
\usepackage{amstext}
\usepackage{amsmath}
\usepackage{amssymb}
\usepackage{verbatim}
\usepackage{psfrag}
\usepackage{dcolumn}

\newcommand{\ud}{\mathrm{d}}
\newcommand{\bra}[1]{\langle #1|}
\newcommand{\ket}[1]{| #1\rangle}
\newcommand{\vect}[1]{\boldsymbol{#1}}
\newcommand{\eq}[1]{Eq.~\eqref{#1}}
\newcommand{\fig}[1]{Fig.~\ref{#1}}
\newcommand{\stn}[1]{Sec.~\ref{#1}}
\newcommand{\be}{\begin{equation}}
\newcommand{\ee}{\end{equation}}
\newcommand{\ti}[1]{\text{#1}}
\newcommand{\mc}[1]{\mathcal{#1}}
\newcommand{\w}{\omega}
\newcommand{\mean}[1]{\langle #1 \rangle}

\begin{document}

\title{Femtosecond dynamics and laser control of charge transport in \emph{trans}-polyacetylene}
\author{Ignacio Franco}
%\email[]{ifranco@chem.utoronto.ca}
\affiliation{Chemical Physics Theory
Group, Department of Chemistry, and  Center for Quantum Information and
Quantum Control, University of Toronto, Toronto, Ontario, Canada.}
\author{Moshe Shapiro}
\affiliation{Chemical Physics Department, The Weizmann Institute,
Rehovot, Israel, and Departments of Chemistry and Physics, The University of
British Columbia, Vancouver, B.C., Canada}
\author{Paul Brumer}
%\email[]{pbrumer@chem.utoronto.ca}
\affiliation{Chemical Physics Theory
Group, Department of Chemistry, and  Center for Quantum Information and
Quantum Control, University of Toronto, Toronto, Ontario, Canada.}

\date{\today}
%----------------------------------------------------------------%
\begin{abstract}
The induction of dc electronic transport in rigid and flexible 
\emph{trans}-polyacetylene oligomers according to the $\w$ 
vs.~$2\w$ coherent control scenario is investigated using  a 
quantum-classical mean field approximation. The approach involves 
running a large ensemble of mixed quantum-classical trajectories 
under the influence of $\w+2\w$ laser fields, and choosing the 
initial conditions by sampling the ground-state Wigner 
distribution function for the nuclei. The vibronic couplings are 
shown to change the mean single-particle spectrum,  introduce 
ultrafast decoherence, and enhance intramolecular vibrational and 
electronic relaxation. Nevertheless, even in the presence of 
significant couplings, limited coherent control of the electronic 
dynamics is still viable, the most promising route involving the 
use of fs pulses with a duration that is comparable to  the 
electronic dephasing time. The simulations offer a realistic 
description of the behavior of a simple coherent control scenario 
in a complex system, and provide a detailed account of the 
femtosecond photoinduced vibronic dynamics of a conjugated 
polymer.
\end{abstract}
%----------------------------------------------------------------%
\maketitle

%----------------------------------------------------------------%
\section{Introduction}

Conjugated polymers are of interest for their broad technological
applications~\cite{HeegerAJ:NobLSm, FriendRH:Elecp, Forrest} and because they
are model systems that offer insight into  the properties of
soft organic and biological matter.  As almost every photochemical,
photophysical, spectroscopic and charge and energy transfer process in these
materials involves dynamics of photoexcited states, the possibility to
manipulate, at a molecular level, the dynamical properties of these
excitations by means of lasers with well defined phases may have  profound
technological implications.

As a first step toward this  goal, we are interested in 
manipulating the dynamics of photoexcited electrons along the 
backbone of a conjugated polymer using the principles of coherent 
control~\cite{paul}. Specifically, we want to induce net dipoles 
along the material  without introducing a bias voltage. For this 
we apply zero-mean  laser pulses with frequency components $\w$ 
and $2\w$. Such fields are known to induce phase-controllable 
dipoles or currents in anharmonic symmetric systems~\cite{ignacionew} even when 
they have a zero-temporal mean, a phenomenon that is referred to 
as laser-induced symmetry breaking. This rectification effect 
first appears in the third order nonlinear response of the system 
to the radiation field~\cite{francoprl}. At this order the system 
mixes the frequencies and harmonics of the $\w+2\w$ field, 
generating a phase-controllable zero harmonic (DC) component in 
the response. The DC component in the  photoinduced dipoles  
is typically of the form $\overline{\langle \mu \rangle} \sim \epsilon_\w^2 \epsilon_{2\w} \cos(\phi_{2\w} - 2\phi_\w)$, where $\epsilon_{n\w}$ and $\phi_{n\w}$ denote the amplitude and phase of the $n\w$ component of the field. Hence, simply  by varying the relative phase between the two incident lasers one can exert control over the magnitude and sign of the symmetry breaking. 

Laser-induced symmetry breaking~\cite{paul, francoprl} has been 
demonstrated in a wide variety of systems ranging from atoms to 
solid state samples. Experimentally, it has been implemented for 
generating anisotropy in  atomic
photoionization~\cite{yin_1vs2}, photocurrents in quantum 
wells~\cite{dupont_1vs2}, intrinsic 
semiconductors~\cite{hache_1vs2} and metal surfaces~\cite{Gudde}, as well as directed diffusion 
in symmetric optical lattices~\cite{schiavoni}. Theoretically, it 
has been studied for generating transport in 
doped~\cite{paul_1vs2} and bulk semiconductors through 
interband~\cite{atanasov_1vs2} and intraband~\cite{pronin_1vs2} 
excitations, in graphene and carbon nanosheets~\cite{mele_1vs2} 
and in molecular wires~\cite{prlwire, hanggi_1vs2, lehmann_1vs2}, 
among others. The setup is of interest since, with current laser 
technology, it can be employed to generate controlled transport 
on a femtosecond timescale.

Of major concern when using  lasers to generate electronic 
transport   is  the influence of the lattice dynamics on the 
rectification.  The lattice can induce ultrafast 
dephasing~\cite{breuer,weiss} processes that have deleterious 
effects on the control. Only a few attempts to quantify this 
effect  exist. In all of them, the explicit dependence of the 
dynamics  on  the nuclear degrees of freedom is eliminated, and 
the  effect of the lattice has been  modeled through  
phenomenological relaxation~\cite{hache_1vs2}, stochastic 
forces~\cite{pronin_1vs2} or thermal baths~\cite{lehmann_1vs2}. 
Although formulations of this type have a considerable domain of 
applicability, they are not appropriate here since   the 
electronic dynamics and spectroscopic observables in conjugated 
polymers depend  on the detailed dynamics of the molecular 
backbone~\cite{sergei2, sergei, francojacs}.  This  
coupling between, and mutual influence of, electronic and 
vibrational degrees of freedom gives rise to the very rich 
photophysics of solitons, polarons and  breathers and  
constitutes an important distinction  between ``soft'' materials 
and rigid solids~\cite{HeegerAJ:NobLSm, SSH,sergei}. 
Furthermore,  the  constant exchange of energy between the 
electronic and nuclear degrees of freedom during 
photoexcitation   keeps the vibrations  out of equilibrium, 
rendering the thermal description inappropriate.

Here we numerically investigate the effect of vibronic couplings on the
applicability of laser-induced symmetry breaking in $\pi$-conjugated systems
by explicitly following the dynamics of both electronic and vibrational
degrees of freedom.  The simulations presented below provide a realistic
description of the behavior of a simple laser control scenario in a
complex system, and offer a detailed account of the femtosecond vibronic
dynamics of a conjugated polymer.

As a minimal microscopic  model for conjugated polymers we adopt 
the Su-Schrieffer-Heeger (SSH) Hamiltonian for 
\emph{trans}-polyacetylene (PA)~\cite{HeegerAJ:NobLSm, SSH}. The 
SSH model treats the polymer chain in terms of a one-dimensional 
tight-binding model in which the $\pi$ electrons are coupled to 
distortions in the polymer backbone by electron-vibrational 
interactions. It neglects quasiparticles interactions, assuming 
that they are relatively weak due to screening. Despite of its 
simplicity, the SSH  Hamiltonian has been remarkably successful 
in reproducing the band structure and the dynamics of excitations 
in PA. With it, we follow numerically the highly nonlinear 
coupled dynamics of electronic and vibrational degrees of freedom 
in neutral PA chains during and after photoexcitation with $\w 
+2\w$  laser pulses of varying frequency, width and intensity. 
The simulations are performed in a mean-field (Ehrenfest) 
approximation~\cite{tully,halcomb,bornemann,prezdho} in which the 
nuclei are treated classically and the electrons quantum 
mechanically. Mean-field 
dynamics is the 
simplest mixed quantum-classical method that allows transfer of 
energy between quantum and classical coordinates with a proper 
conservation of energy~\cite{tully,halcomb,bornemann,prezdho}, and where transitions between 
instantaneous eigenstates are allowed.   Allowing for change in 
the occupation of the electronic levels is crucial because the 
laser is constantly inducing electronic transitions and, 
additionally, electronic levels can approach one another closely  
during the dynamics and may lead to nonadiabatic transitions 
between electronic states. It is also a
tractable method, as shown below, for
treating the vibronic dynamics of multiple electronic states in the presence
of a laser.
 
 In order to  incorporate the effects of lattice fluctuations on the dynamics 
we follow the evolution of an  ensemble of quantum-classical 
trajectories. The initial conditions are obtained by using 
importance sampling for the nuclear Wigner phase space 
distribution in the harmonic approximation. In this way, the 
resulting  averaged dynamical observables   retain relevant 
correlations between the electronic and vibrational degrees of 
freedom and are subject to electron-vibrational induced 
decoherence and relaxation. 

This analysis  begins by briefly describing the model  
Hamiltonian  (\stn{PAsec:SSH}) and deriving the equations of 
motion for the vibronic  dynamics (\stn{PAsec:dynamics}). Then we 
turn to the important problem of choosing initial conditions for 
the evolution  (\stn{PAsec:initial}) and we introduce the 
geometric and spectroscopic observables that are used to monitor 
the dynamics (\stn{PAsec:observables}). Our main results are 
presented  in \stn{PAsec:results}. Specifically, after discussing 
the properties of the initial state 
(\stn{PAsec:initialresults}),  we investigate the typical 
dynamics of the chain generated by short (10 fs) and long (300 
fs) pulses, and estimate the typical electronic dephasing times 
in PA oligomers (\stn{PAsec:dynamicsresults}). Then, we turn our 
attention to the effects that the lattice dynamics has on 
laser-induced symmetry breaking (\stn{PAsec:control}).  The main 
results are summarized in \stn{PAsec:conclusions}.

%----------------------------------------------------------------%
\section{Model and methods}
\label{PAsec:methodology}

%----------------------------------------------------------------%
\subsection{The SSH model}
\label{PAsec:SSH}

The SSH Hamiltonian~\cite{HeegerAJ:NobLSm, SSH} models the PA oligomer  
as a one-dimensional tight-binding  chain, each site representing 
a CH unit. The Hamiltonian for an $N$-membered chain has the  form:
%----------------------------------------------------------------%
\begin{equation}\label{PAeq:SSH}
H_{\textrm{SSH}} = H_{\pi} + H_{\pi-\textrm{ph}} + H_{\textrm{ph}},
\end{equation}
%----------------------------------------------------------------%
where
%----------------------------------------------------------------%
\begin{equation}
H_{\pi} = -t_0 \sum_{n=1}^{N-1} \sum_{s=\pm 1}
(c_{n+1,s}^{\dagger}c_{n,s} + c_{n,s}^{\dagger}c_{n+1,s}),
\end{equation}
%----------------------------------------------------------------%
describes the hopping of $\pi$ electrons along the chain without spin flip
characterized by the lowest-order hopping integral $t_0$. The operator $c_{n,s}^{\dagger}$
($c_{n,s}$)  creates (annihilates) a fermion in site $n$ with spin $s$ and
satisfies the usual fermionic anticommutation relations
$\{c_{n,s},c_{m,s'}^{\dagger}\}=\delta_{n,m}\delta_{s,s'}$.  The
$\pi$-electron-ion interaction term is given by
%----------------------------------------------------------------%
\begin{equation}
H_{\pi-\textrm{ph}} = \alpha\sum_{n=1}^{N-1}\sum_{s= \pm 1}
\left(u_{n+1} - u_{n}\right)
(c_{n+1,s}^{\dagger}c_{n,s} + c_{n,s}^{\dagger}c_{n+1,s}),
\end{equation}
%----------------------------------------------------------------%
where $u_n$ is the displacement of the $n$-th site in the $x$ 
direction from the perfectly periodic position $x=na$, with $a$ 
as the lattice constant. The operator $H_{\pi-\rm{ph}}$  couples 
the electronic states to the molecular geometry and provides a 
first-order correction to the hopping integral with $\alpha$ as 
the coupling constant. Finally, the nuclear Hamiltonian is taken 
to be
%----------------------------------------------------------------%
\begin{equation}
H_{\textrm{ph}} = \sum_{n=1}^{N} \frac{p_n^2}{2M} +
\frac{K}{2}\sum_{n=1}^{N-1} \left(u_{n+1} - u_{n}\right)^{2},
\end{equation}
%----------------------------------------------------------------%
where $M$ is the mass of the CH group, $p_n$ is the momentum 
conjugate to $u_n$ and $K$ is an effective spring constant.

The SSH model is an effective~\cite{SSH}  empirical model for non-interacting
quasiparticles in PA.  The electron-vibrational coupling and the hopping integral
can  be viewed as parameters in which screening and other high-energy effects
have already been taken into account. The effect of the residual interactions
that cannot be accounted for by a simple renormalization of the one-electron
Hamiltonian are completely neglected. 

Throughout this work, we use the standard SSH parameters for 
PA~\cite{SSH}: $\alpha = 4.1$ eV/\AA, $K=21$ eV/\AA$^2$, 
$t_0=2.5$ eV,  $M=1349.14$ eV fs$^2$/\AA$^2$ and $a=1.22$ \AA. 
Results using this set of parameters agrees qualitatively well 
with experimentally determined properties. 

%Alternative sets have 
%been constructed from \emph{ab initio} 
%computations~\cite{parameters}.  

%----------------------------------------------------------------%
\subsection{Photoinduced dynamics}
\label{PAsec:dynamics}

To simulate excitation of the chain by a femtosecond laser pulse, 
an  interaction with an electric field  in dipole approximation 
is added to the SSH Hamiltonian. The electric field is taken into 
account to all orders and the  dynamics is followed in the 
mean-field (Ehrenfest) approximation.  
Our approach resembles that of Johansson \emph{et 
al.}~\cite{johansson1, johansson2} and 
Streitwolf~\cite{streitwolf1}, but  we  dynamically propagate  an 
\emph{ensemble} of quantum-classical trajectories rather than a 
single trajectory that starts from the optimal geometry. 
Single-trajectory  approaches~\cite{sergei, sergei2, francojacs, 
johansson1, johansson2,streitwolf1, sergeibreather}
 offer valuable insights into the time scales involved in the
vibronic dynamics, but are insufficient for our purposes because 
including the decoherence effects requires averaging over 
the nuclear degrees of freedom. Further, during this study we 
have observed that the vibronic dynamics of the SSH chain tends 
to exhibit deterministic chaos and, hence, individual 
trajectories become inaccurate for sufficiently long integration 
times. By contrast, results for an ensemble of trajectories are 
meaningful even for long integration times, provided the 
shadowing theorem~\cite{shadowing1, shadowing3} holds.

In the presence of a radiation field, the total Hamiltonian of the system
assumes the form
%----------------------------------------------------------------%
\be
\label{PAeq:totham}
H_\ti{S}(t) = H_{\text{SSH}} + H_{E}(t),
\ee
%----------------------------------------------------------------%
where $ H_{E}(t)$  describes the interaction of the chain with an
external electric field $E(t)$  in the dipole approximation and Coulomb
gauge,
%----------------------------------------------------------------%
\begin{equation}
H_{E}(t) = -(\mu_e + \mu_i)E(t).
\end{equation}
%----------------------------------------------------------------%
Here $\mu_e= - |e| \sum_{n,s} x_n c_{n,s}^{\dagger}c_{n,s}$ and $\mu_i=+ |e| \sum_{n} x_n$ are the electronic and ionic dipole moments, 
 $x_n = na + u_n$ is the monomer position operator, $a$  the 
lattice parameter and $-|e|$ the unit electronic charge.   The 
Coulomb gauge is consistent with our open chain boundaries.  As 
a  field  we use the  femtosecond $\w+2\w$ Gaussian pulses 
detailed in Table~\ref{tbl:laserparameters} at a variety of 
frequencies noted in the text.

%----------------------------------------------------------------%
\begin{table}
\caption{\label{tbl:laserparameters} Parameters and labels defining the
femtosecond laser pulses used
$E(t) = \exp(- (t-T_c)^2/T_\ti{W}^2)(\epsilon_\w \cos(\w t +
\phi_\w) + \epsilon_{2\w} \cos(2\w t + \phi_{2\w}))$. Here $I_{2\w}$ is the intensity of the $2\w$ component at maximum field strength.}
\centering
\begin{tabular}{|c| c| c | c | c| c|}
\hline
\hline
Label & $T_c$ (fs) & $T_\ti{W}$ (fs) &
$\epsilon_{2\w}$ (V \AA$^{-1}$) & $\epsilon_{\w}/\epsilon_{2\w}$  &
 $I_{2\w}$ (W cm$^{-2}$) \\
\hline
 \textbf{f1} & 900 & 300 & $8.70\times10^{-3}$ & 2.82 & $1.0\times 10^9$ \\
\textbf{f2} & 900 & 300 & $4.00\times10^{-2}$ & 2.82 & $2.1\times 10^{10}$ \\
\textbf{f3} & 50 & 10 & $8.70\times10^{-3}$ & 2.82 & $1.0\times 10^9$ \\
 \textbf{f4} & 50 & 10 & $4.00\times10^{-2}$ & 2.82 & $2.1\times 10^{10}$ \\
\hline
\hline
\end{tabular}
\end{table}
%----------------------------------------------------------------%

In the mean-field approximation
the nuclei move classically on a mean-field potential energy 
surface with forces given by~\cite{tully,halcomb,bornemann,prezdho}
%----------------------------------------------------------------%
\begin{equation}
\label{PAeq:mf1}
 \dot{p}_{n}  = - \langle \Psi(t)| \frac{\partial H_\ti{S}(t)}
{\partial u_n}|\Psi(t)\rangle,
\end{equation}
%----------------------------------------------------------------%
where $\ket{\Psi (t)}$ denote the antisymmetrized 
$\mc{N}$-electron wave function. 
The electrons are assumed to respond instantaneously to the nuclear motion and,
hence, $\ket{\Psi(t)}$ satisfies the time-dependent Schrodinger equation
%----------------------------------------------------------------%
\begin{equation}
\label{PAeq:mf2}
i\hbar \frac{\partial}{\partial t} |\Psi( t)\rangle =
H_{\text{elec}}(\vect{u}(t), t)|\Psi( t)\rangle,
\end{equation}
%----------------------------------------------------------------%
where $H_\text{elec}$ contains the  electronic and the 
electron-nuclei interaction terms of the total Hamiltonian, and $\vect{u}= (u_1, u_2, \cdots, u_N)$.   Equations~\eqref{PAeq:mf1} 
and~\eqref{PAeq:mf2} define the Ehrenfest method. Feedback 
between the fast and slow degrees of freedom is incorporated in 
both directions in an average self-consistent way.  Note that 
the  mean-field approximation avoids the expansion of the 
electronic wave function in terms of adiabatic basis functions. 
Equations~\eqref{PAeq:mf1} and \eqref{PAeq:mf2} can therefore be 
integrated directly, making the implementation of the method 
particularly simple.

The electronic part of  the total system Hamiltonian [\eq{PAeq:totham}] reads
%----------------------------------------------------------------%
\be
\begin{split}
\label{PAeq:elecham}
H_\ti{elec}(t)  = \sum_{n=1,s}^{N-1} [-t_0+\alpha(u_{n+1}-u_n)]
(c_{n+1,s}^\dagger c_{n,s} \\
+ c_{n,s}^\dagger c_{n+1,s}) + |e| \sum_{n=1,s}^{N} x_{n} c_{n,s}^\dagger
c_{n,s} E(t).
\end{split}
\ee
%----------------------------------------------------------------%
Since $H_\ti{elec}(t)$  is a single particle operator, the electronic properties
of the system are completely characterized by the  electronic reduced density 
matrix, defined by
%----------------------------------------------------------------%
\begin{equation}
\label{PAeq:rho}
\rho_{n,m}(t) =\sum_s \langle \Psi (t)| c_{n,s}^{\dagger} c_{m,s}
| \Psi(t) \rangle.
\end{equation}
%----------------------------------------------------------------%
From \eq{PAeq:mf2} it follows that the dynamics of $\rho_{n,m}(t)$ is governed by
%----------------------------------------------------------------%
\be
\label{eq:rdmeom}
\begin{split}
i \hbar \frac{\ud}{\ud t} \rho_{n,m}(t) & =
\sum_{s} \bra{\Psi(t)}[ c_{n,s}^\dagger c_{m,s}, H_\ti{elec}(t)] \ket{\Psi(t)}  \\
& = \sum_{m'}\left(h_{m,m'}(t)  \rho_{n,m'}(t) - h_{m',n}(t) \rho_{m',m}(t)\right),
\end{split}
\ee
%----------------------------------------------------------------%
where $h_{n,m}(t) = \bra{n,s} H_\ti{elec}(t)\ket{m,s}$ are the single-particle matrix elements of $H_\ti{elec}(t)$, and $\ket{n,s}=c_{n,s}^\dagger\ket{0}$ where $\ket{0}$ is the vacuum state. In order to integrate  this equation it is useful to 
employ an orbital decomposition for $\rho_{n,m}(t)$, as described below.

Let $\ket{\epsilon,s}$ be the eigenorbitals of the system at preparation time, defined by the eigenvalue relation $H_\ti{elec}(t=0) \ket{\epsilon,s} = \epsilon \ket{\epsilon,s}$.
Using this basis,  the initial electronic reduced density matrix  can be expressed as
%----------------------------------------------------------------%
\be
\begin{split}
\label{eq:rdminitial}
\rho_{n,m}(0)
= \sum_{\epsilon, \epsilon'=1}^{N}\sum_{s} \bra{\epsilon,s}n,s\rangle \langle m,s\ket{\epsilon',s} \\ \bra{\Psi(0)} c_{\epsilon,s}^\dagger c_{\epsilon',s} \ket{\Psi(0)},
\end{split}
\ee
%----------------------------------------------------------------%
where $c_{\epsilon,s}^\dagger$ creates a fermion with spin $s$ in the orbital with energy $\epsilon$ (i.e. $\ket{\epsilon,s}= c_{\epsilon,s}^\dagger\ket{0}$), and $\bra{\Psi(0)} c_{\epsilon,s}^\dagger c_{\epsilon',s} \ket{\Psi(0)}$ characterizes the initial electronic distribution among the single particle states. In  writing  \eq{eq:rdminitial}  we have employed the basis transformation function $ c_{n,s}  = \sum_{\epsilon=1}^{N} \bra{n,s} \epsilon,s\rangle c_{\epsilon,s}$, and its hermitian conjugate, which takes into account the fact that $\bra{\epsilon,s} n,s'\rangle= \bra{\epsilon,s} n,s\rangle\delta_{s,s'}$ for the Hamiltonian under consideration.  Upon time evolution  we assume that $\rho_{n,m}(t)$ maintains the form in \eq{eq:rdminitial}. That is,
%----------------------------------------------------------------%
\be
\begin{split}
\label{eq:rdmtime}
\rho_{n,m}(t) =
\sum_{\epsilon, \epsilon'=1}^{N}\sum_{s} \bra{\epsilon(t),s}n,s\rangle \langle m,s\ket{\epsilon'(t),s} \times \\ \bra{\Psi(0)} c_{\epsilon,s}^\dagger c_{\epsilon',s} \ket{\Psi(0)}.
\end{split}
\ee
%----------------------------------------------------------------%
The utility of this ansatz  is that if the time-dependent
orbitals $\ket{\epsilon(t),s}$ satisfy the single-particle Schrodinger equation
%----------------------------------------------------------------%
\be
\label{PAeq:electmp}
 i\hbar \frac{\ud}{\ud{t}}\ket{\epsilon(t),s}
=   H_\ti{elec}(t)\ket{\epsilon(t),s},
\ee
%----------------------------------------------------------------%
with initial conditions $ \ket{\epsilon(t=0),s} =  
\ket{\epsilon,s}$, the reduced density matrix  automatically 
satisfies the correct equation of motion [\eq{eq:rdmeom}].

With the exception of the simulations presented in 
\stn{PAsec:decoherence}, the initial electronic state 
$\ket{\Psi(0)}$ is taken to be  a single Slater determinant for 
which
%----------------------------------------------------------------%
\be
\label{PAeq:distrib}
\bra{\Psi(0)} c_{\epsilon,s}^\dagger c_{\epsilon',s} \ket{\Psi(0)}= \delta_{\epsilon, \epsilon'} f(\epsilon,s),
\ee
%----------------------------------------------------------------%
where $f(\epsilon,s)$ is the  initial distribution function that 
takes values $0$ or $1$ depending on the initial occupation of 
each level with energy $\epsilon$ and spin $s$. In this case, 
$\rho_{n,m}(t)$ assumes the simplified form
%----------------------------------------------------------------%
\begin{equation}
\label{PAeq:rdmsimple}
\rho_{n,m}(t) = \sum_{\epsilon=1}^{N} \sum_{s}\bra{\epsilon(t),s} n,s\rangle \langle m,s\ket{\epsilon(t),s} f(\epsilon,s).
\end{equation}
%----------------------------------------------------------------%
Henceforth we drop spin labels since orbitals with 
opposite spin satisfy the identical equation of motion.

Given Eqs.~\eqref{PAeq:mf1} and \eqref{PAeq:rho} and the total Hamiltonian of the
system~\eq{PAeq:totham}, the equations for the nuclear trajectories are
given as
%----------------------------------------------------------------%
\begin{equation}
\label{PAeq:nuclei}
\begin{split}
\dot{u}_n(t)  =& \frac{p_n(t)}{M}; \\
\dot{p}_n(t)  =&  - K\left(2u_n(t) - u_{n+1}(t) - u_{n-1}(t)\right) \\
&+ 2\alpha\textrm{Re}\left\{ \rho_{n, n+1}(t)
-   \rho_{n, n-1}(t) \right\} \\
&- |e|E(t) \left( \rho_{n,n}(t) -1\right).
\end{split}
\end{equation}
%----------------------------------------------------------------%
 The chain is taken to be clamped so that $u_1(t)=u_N(t)=0$ and
$p_1(t)=p_N(t)=0$ for all time, and \eq{PAeq:nuclei}  is valid for
$n=2, \cdots, N-1$. In turn, the orbitals that form 
$\rho_{nm}(t)$ satisfy \eq{PAeq:electmp}, so that
%----------------------------------------------------------------%
\be
\label{PAeq:elec}
\begin{split}
 i\hbar \frac{\ud}{\ud t} \langle n \ket{\epsilon(t)}
 = & \left[-t_0 + \alpha (u_{n+1}(t)-u_n(t))\right]\langle n+1\ket{\epsilon(t)} \\
+ & \left[-t_0 + \alpha (u_{n}(t)-u_{n-1}(t))\right]\langle{ n-1}\ket{\epsilon(t)} \\
 + &  |e|E(t) \left(na+u_n(t)\right) \langle n\ket{\epsilon(t)},
\end{split}
\ee
%----------------------------------------------------------------%
for $n, \epsilon=1, \cdots, N$. Since the electrons are confined 
within the chain,  $\bra{n}\epsilon(t)\rangle=0$ for $n \notin 
\{1, \cdots, N\}$. Equations~\eqref{PAeq:nuclei} 
and~\eqref{PAeq:elec} constitute a closed set of $N(N+2)$  
coupled first-order differential equations, which are integrated 
using the eighth-order  Runge-Kutta method  with step-size 
control~\cite{rksuite}.

%----------------------------------------------------------------%
\subsection{The initial conditions}
 \label{PAsec:initial}

A crucial aspect of the problem is the choice of initial 
conditions.  The oligomer is   assumed to be in the ground electron-vibrational state. 
Hence, we first determine the optimal geometry of an $N$-membered oligomer with open boundaries and clamped ends 
($u_1 = u_N = 0$)  by an iterative 
self-consistent procedure,  and then  perform a normal mode analysis in the 
ground electronic surface as detailed in Ref.~\onlinecite{phonon1}.   This procedure provides
the nuclear ground-state wave function in the harmonic approximation.  
A phase-space like 
description of the resulting nuclear quantum state is obtained  by constructing the associated  nuclear Wigner phase-space
distribution~\cite{ wigner,  tatarskii} function $\rho_\ti{W}(\vect{u}, \vect{p})$,  which   is
just the product of the Wigner distributions associated with each vibrational
mode
%----------------------------------------------------------------%
\be
\label{PAeq:wignerlattice}
\rho_\ti{W}(\vect{u}, \vect{p}) = \prod_{j=1}^{N-2} \rho_j(Q_j(\vect{u}),
P_j(\vect{p})).
\ee
%----------------------------------------------------------------%
Here  $Q_j(\vect{u})$ is the normal mode coordinate of the $j$-th mode and
$P_j(\vect{p})$ its  conjugate momentum. In the ground state, the Wigner distribution of each 
normal mode is given by~\cite{wigner}
%----------------------------------------------------------------%
\be
\rho_j(Q_j, P_j) =\frac{1}{\pi\hbar} \exp({- M \w_j Q_j^2/\hbar})
\exp({-P_j^2/\hbar\w_j M}),
\ee
%----------------------------------------------------------------%
where $\w_j$ is the frequency of the $j$th  mode. The $2N-4$ dimensional phase-space distribution in
\eq{PAeq:wignerlattice} completely characterizes the initial quantum state of
the nuclei.

The ensemble of lattice initial conditions,
$\{\vect{u}^i(0), \vect{p}^i(0)\},$
for the quantum-classical dynamics is obtained from a
Monte Carlo  sampling of the
nuclear Wigner phase space distribution of \eq{PAeq:wignerlattice}.
The average classical energy of the resulting ensemble
coincides numerically with the zero-point energy of the lattice. 
 The associated  initial values for the orbitals 
$\{\langle n \ket{\epsilon^i(0)}\}$  are  obtained by 
diagonalizing the electronic part of the  Hamiltonian in the 
initial lattice geometries $\{\vect{u}^i(0)\}$.  Each  initial 
condition $i$, together with the equations of 
motion~\eqref{PAeq:nuclei} and~\eqref{PAeq:elec},  defines a 
quantum-classical trajectory $(\vect{u}^i(0), \vect{p}^i(0), 
\ket{\Psi^i(0)}) \to (\vect{u}^i(t), \vect{p}^i(t), 
\ket{\Psi^i(t)})$. The set is propagated using a parallel algorithm and  used to obtain ensemble averages.
 In this 
manner, the dynamics reflects the effects of lattice fluctuations 
and  the initial quantum phase-space distribution of the nuclei.

Since laser-induced symmetry breaking is an effect that depends 
on the third-order response of the system to the field,  adequate 
convergence of the results with the number of initial conditions 
requires a large number of initial configurations, 
$\mc{O}(10^3\times(2N-4))$.   This is the main computational 
bottleneck of the present approach and  limits our analysis to 
modestly sized oligomer chains.

%----------------------------------------------------------------%
\subsection{Dynamical observables}
\label{PAsec:observables}

During the vibronic dynamics induced by the laser field  the
molecular geometry  distorts and this, in turn, induces strong 
changes in the electronic wave function.  Below we describe the 
geometric and spectroscopic observables that we use to follow  
the evolution of both nuclei and electrons.

Geometrical changes in the polymer backbone are  characterized  by the bond
length alternation parameter $r_n$, which  quantifies the
homogeneity in the distribution of $\pi$ electrons over the bonds. The ensemble
average of this quantity  is defined by
%----------------------------------------------------------------%
\be
\label{PAeq:bla}
\begin{split}
\mean{r_n} &= \frac{1}{\mc{M}}
\sum_{i=1}^{\mc{M}}\frac{(-1)^n}{2}(l_{n-1}^i - l_{n}^i)\\ 
& = \frac{1}{\mc{M}}\sum_{i=1}^{\mc{M}}
\frac{(-1)^n}{2}\left(2u_n^i - u_{n-1}^i - u_{n+1}^i\right),
\end{split}
\ee
%----------------------------------------------------------------%
where $l_n^i$ is the bond length between the $n$-th and $(n+1)$-th atoms along
the chain in the $i$-th member of the ensemble and $\mc{M}$ is the cardinality
of the ensemble. When the alternation between  single and double bonds is
perfect, the bond length alternation is constant (apart from end effects) and
takes a value of $r_0\approx 0.08$ \AA. An enhancement in the electronic
delocalization tends to equalize the bond lengths in the polymer resulting in
$|r_n|/r_0 \ll 1$.

The state of the electronic degrees of freedom is  completely characterized by
the  ensemble-averaged electronic reduced density matrix,
%----------------------------------------------------------------%
\be
\label{PAeq:rdm}
\overline{\rho}_{n,m}(t)  = \frac{1}{\mc{M}}\sum_{i=1}^{\mc{M}} \rho_{n,m}^i(t)
\ee
%----------------------------------------------------------------%
where   $\rho_{n,m}^i$ is defined by \eq{PAeq:rho}.  The diagonal elements of
$\overline{\rho}_{n,m}(t)$ represent the density of charge along the chain.
The off-diagonal elements represent the electronic coherences between different
sites.

A compact description of the electronic dynamics is offered by the polarization
of the chain, defined by
%----------------------------------------------------------------%
\be
\label{PAeq:x}
\langle \mu (t) \rangle =  \frac{|e|}{\mc{M}}\sum_{i=1}^\mc{M}
\sum_{n=1}^{N} x_n^i(t)(1-\rho_{n,n}^i),
\ee
%----------------------------------------------------------------%
where $x_n^i(t) = (na+u_n^i(t))$ is the position of site $n$  at time $t$ in
the $i$-th trajectory.  The first term in \eq{PAeq:x} comes from the dipole due  to the nuclei, while the second one quantifies the electronic contributions.
Any laser-induced symmetry breaking  must manifest  as a DC component  in
$\langle \mu (t)\rangle$. In order to isolate the zero-frequency component from
the oscillatory terms  we integrate the signal over time
%----------------------------------------------------------------%
\be
\label{PAeq:cumx}
\mean{C(t)} = \frac{1}{L T_\ti{W}}\int_0^t \ud t' \mean{\mu(t')}.
\ee
%----------------------------------------------------------------%
This gives  the net cumulative dipole induced by the field,
weighted over
the chain length $L$ and the temporal pulse width $T_\ti{W}$.

For each member of the ensemble the instantaneous eigenorbitals of the
electronic
Hamiltonian (including the field) are defined by the eigenvalue relation
%----------------------------------------------------------------%
\be
H_\ti{elec}^i(t)\ket{\gamma^i(t)}
 = \epsilon_{\gamma}^i(t) \ket{\gamma^i(t)},
\ee
%----------------------------------------------------------------%
where $H_\ti{elec}^i(t)$ is the electronic Hamiltonian for 
trajectory $i$ at time $t$.  At each instant in time the 
instantaneous eigenorbitals for different trajectories  differ 
since each trajectory experiences a different vibronic evolution. 
For each trajectory, at a given $t$,  the orbitals 
$|\epsilon^i(t)\rangle$ satisfying Eq.~\eqref{PAeq:elec} may be 
expanded as a linear combination of the instantaneous 
eigenfunctions $|\gamma^i\rangle$, $ |\epsilon^i (t)\rangle = 
\sum_\gamma a_{\gamma, \epsilon}^i(t)  | \gamma^i (t)\rangle$, 
where $ a_{\gamma, \epsilon}^i(t) = 
\langle\gamma^i(t)|\epsilon^i(t)\rangle$. The occupation number  
$n_\gamma^i$ of the instantaneous states is thus  given by
%----------------------------------------------------------------%
\begin{equation}\label{PAeq:occup}
n_\gamma^i(t) = \sum_{\epsilon, s} | \langle \gamma^i (t) | \epsilon^i(t)
\rangle|^2 f(\epsilon, s) =   \sum_{\epsilon, s} |a_{\gamma, \epsilon}^i|^2
f(\epsilon, s),
\end{equation}
%----------------------------------------------------------------%
and varies as the system evolves. This contrasts with purely adiabatic
dynamics with fixed level occupation. For the ensemble, we define the mean
occupation and mean energy of the instantaneous eigenstates as
%----------------------------------------------------------------%
\be
\begin{split}
\langle \epsilon_\gamma(t) \rangle = \frac{1}{\mc{M}} \sum_{i=1}^{\mc{M}}
\epsilon_\gamma^i(t),~~~~
\langle n_\gamma(t) \rangle  = \frac{1}{\mc{M}} \sum_{i=1}^{\mc{M}}
n_\gamma^i(t),
\end{split}
\ee
%----------------------------------------------------------------%
for $\gamma =1, \cdots, N$.These quantities are physically acceptable if the
ordering of the eigenstates does not vary among trajectories, a property that
we  assume. Throughout the text the label $\gamma$ is employed to number the instantaneous eigenstates of the oligomer in ascending order,  with $\gamma=1, N/2$ and $N/2+1$ denoting the lowest energy, HOMO and LUMO orbitals respectively.

%----------------------------------------------------------------%
\section{Results and discussion}
\label{PAsec:results}

Consider the ability of $\w + 2\w$ fields to induce symmetry 
breaking in PA.  The dynamics is characterized using the 
observables discussed in Sec.~\ref{PAsec:observables} as well as 
by the different contributions to the chain energy. Vibrational effects are made
explicit  by contrasting the vibronic dynamics of the 
oligomers  with the evolution of a single trajectory 
for an equivalent but rigid system initially  in the optimal geometry. The chain 
is made rigid by multiplying the mass of the CH groups  by a 
factor of $10^6$. In this way  the lattice is made to move a 
thousand times more slowly while maintaining the same value of 
the electron-ion interaction.

\subsection{Initial state}
\label{PAsec:initialresults}

We  study neutral oligomers with 20 sites and 20 $\pi$ electrons
($L\approx 23$ \AA) initially in the ground electronic configuration.  This
size was selected  because it reproduces qualitatively well the spectrum and
dimerization pattern of longer chains without making the computational
effort prohibitive.  The initial state for the flexible chain consists of an
ensemble of 100,000 initial conditions obtained by sampling the ground-state
nuclear Wigner distribution (Sec.~\ref{PAsec:initial}).  

\begin{figure}[htbp]
\psfrag{A}[][]{A}
\psfrag{B}{B}
\psfrag{C}{C}
\psfrag{D}{D}
\psfrag{site (m)}[][]{site ($m$)}
\psfrag{site (n)}[][]{site ($n$)}
\psfrag{BLA (A)}[][]{$\mean{r_n}$ (\AA)}
\psfrag{label}[][]{label}
\psfrag{energy (eV)}[][]{energy (eV)}
\centering
\includegraphics[width=0.5\textwidth]{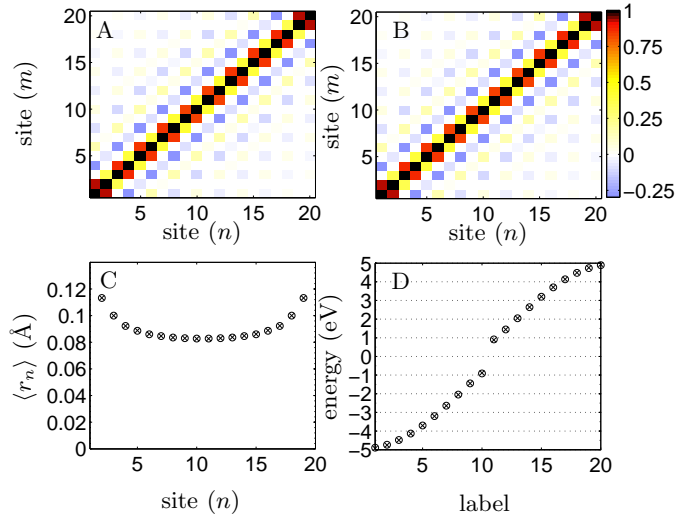}
\caption{Nuclear and electronic properties of a  20 atoms + 20 
$\pi$ electrons SSH-type chain  in the ground-state 
configuration. Properties of the chain in the optimal geometry 
are compared with the average results obtained from an ensemble 
of  100,000 initial configurations. The upper panels show the 
reduced density matrix  (chosen to be real) for (A) the optimal 
geometry and (B) the ensemble of initial states. The color bar is 
given in the far right.  Panel (C): bond length alternation along 
the chain and (D) the single-particle spectrum. ($\circ$) - the 
optimal geometry; ($\times$) - the ensemble average. In both 
cases, the initial average momentum for each atom is zero. }
     \label{fig:initialstate}
\end{figure}

The resulting electronic and geometrical properties of the chain  are shown
in \fig{fig:initialstate}.   The initial geometry of the chain
(\fig{fig:initialstate}C)  consists of a perfect alternation of double and
single bonds, yielding $\mean{r_n}\sim 0.08$ \AA.
The open-ends boundary condition results in a  stronger dimerization
near the edges of the oligomer. The structure is centro-symmetric with the
inversion center residing between sites 10 and 11.
The single-particle spectrum
(\fig{fig:initialstate}D and Table~\ref{tbl:bandwidths})
has a total width of $4t_0=10$ eV. It consists of
$N/2$ fully occupied $\pi$ (valence) states  and $N/2$, initially
empty, $\pi^\star$ (conduction) states, separated by an
energy gap $2\Delta=1.8$ eV. The reduced density matrix
(\fig{fig:initialstate}A-B)  is concentrated along the diagonal of the plot
reflecting the bonding pattern of the oligomer, with the spatial electronic coherence
(measured by the magnitude of the off-diagonal matrix elements)
being $\sim 10-15$ units.

%----------------------------------------------------------------%
%----------------------------------------------------------------%

%----------------------------------------------------------------%
\begin{table}
\caption{\label{tbl:bandwidths} Initial single particle spectrum for the 20 site neutral PA oligomer. The energies $\epsilon_i^0$ were obtained at the optimal geometry,
while $\mean{\epsilon_i}$ denote the average orbital energies for the ensemble.
The ensemble of orbital energies form energy bands distributed
in a Gaussian manner about the band centers,
$f(\epsilon_i) = \frac{1}{\sqrt{2\pi}\sigma_i}
\exp\left({-(\epsilon-\mean{\epsilon_i})^2/2\sigma_i^2}\right)$,  where
$\sigma_i$ is the standard deviation.}
\centering
\begin{tabular}{|c | c |c| c| c | c |c| c|}
\hline
\hline
$i$ & $\epsilon_i^0$ (eV) & $\mean{\epsilon_i}$(eV) & $\sigma_i$ (eV) &
$i$ & $\epsilon_i^0$ (eV) & $\mean{\epsilon_i}$(eV) & $\sigma_i$ (eV)
\\
\hline
\bf{1}  & -4.893 &  -4.930  &  0.034 & \bf{11} &  0.914 &   0.883  &  0.130\\
\bf{2}  & -4.735 &  -4.755  &  0.039 &  \bf{12} &  1.445 &   1.444  &  0.107\\
\bf{3}  & -4.479 &  -4.491  &  0.046 &  \bf{13} &  2.045 &   2.049  &  0.091\\
\bf{4}  & -4.131 &  -4.141  &  0.053 &  \bf{14} &  2.640 &   2.646  &  0.079\\
\bf{5}  & -3.701 &  -3.708  &  0.061 &  \bf{15} &  3.199 &   3.206  &  0.068\\
\bf{6}  & -3.199 &  -3.206  &  0.068 &  \bf{16} &  3.701 &   3.708  &  0.061\\
\bf{7}  & -2.640 &  -2.646  &  0.079 &  \bf{17} &  4.131 &   4.141  &  0.053\\
\bf{8}  & -2.045 &  -2.049  &  0.091 &  \bf{18} &  4.479 &   4.491  &  0.046\\
\bf{9}  & -1.445 &  -1.444  &  0.107 &  \bf{19} &  4.735 &   4.755  &  0.039\\
\bf{10} & -0.914 &  -0.883  &  0.130 &  \bf{20} &  4.893 &   4.930  &  0.034\\
\hline
\hline
\end{tabular}
\end{table}
%----------------------------------------------------------------%

Note that the \emph{average} geometrical and spectroscopic  
properties of the ensemble coincide with the ones obtained from a 
single configuration  in the optimal geometry. However, within 
the ensemble there is a  distribution of values for the 
properties,  displaying important deviations  from the average.  
For example, we find that the ensemble of orbital energies form 
bands of states distributed in a Gaussian fashion about the band 
averages. The banded-Gaussian distribution in energy space is a 
direct result of  the Gaussian distribution of the initial 
geometries. The band-averages $\mean{\epsilon_i}$ and bandwidths 
$\sigma_i$ are enumerated in  Table~\ref{tbl:bandwidths}.  Note 
that the bandwidth of the electronic levels  near the energy gap 
($i=10-11$) is considerably broader than the one displayed by 
states near  the band edges.

Last, due to residual anharmonicities around the equilibrium 
geometry, the distribution in \eq{PAeq:wignerlattice} is not 
completely stationary. Rather, the average values of the initial 
state display some fluctuations under free evolution. These 
fluctuations are very small, introducing only small changes to 
 \fig{fig:initialstate}, of the order of 1\% in the band-gap 
energy and of 3\% in the bond length alternation.

%----------------------------------------------------------------%
\subsection{Rigid vs. flexible chain dynamics}
\label{PAsec:dynamicsresults}

For rigid chains, changes in the electronic spectrum can only 
arise from the Stark shifts induced by the laser field.  By 
contrast, for flexible chains there is a continuous exchange of 
energy between the electronic and nuclear degrees of freedom 
during the photoexcitation process. Below we present two 
illustrative examples of the photoinduced dynamics in PA. First 
we consider the case in which the lattice is photoexcited with a 
10 fs pulse (\stn{PAsec:10fs}). For such short pulses very little 
excited state dynamics occurs during the pulse. In the wake of 
the pulse we observe coherent vibrational breathing motion that 
decays in $\sim 200$ fs due to vibrational energy redistribution. 
We then consider excitation with a 300 fs pulse 
(\stn{PAsec:300fs}). In this case the laser does not generate a 
breathing motion of the lattice. Rather, we observe a gradual 
increase in the electronic delocalization, accompanied by a 
concurrent change in the average spectroscopic and geometrical 
parameters of the chain. 

Last, in \stn{PAsec:decoherence} we quantify the time-scale for 
electronic dephasing for chains of different lengths.

%----------------------------------------------------------------%
\subsubsection{The dynamics  induced by a 10 fs pulse}
\label{PAsec:10fs}

%----------------------------------------------------------------%
\begin{figure}[htbp]
\psfrag{A}[][]{A}
\psfrag{B}[][]{B}
\psfrag{C}[][]{C}
\psfrag{D}[][]{D}
\psfrag{t (fs)}[][]{$t$ (fs)}
\psfrag{helec}[][]{$\mean{ H_\pi + H_{\pi-\text{ph}}}$ (eV)}
\psfrag{hph}[][]{$\mean{ H_\text{ph}}$ (eV)}
\psfrag{he}[][]{$\mean{ H_E}$ (eV)}
\psfrag{h}[][]{$\mean{ H_\ti{S}(t)}$ (eV)}
\centering
\includegraphics[width=0.5\textwidth]{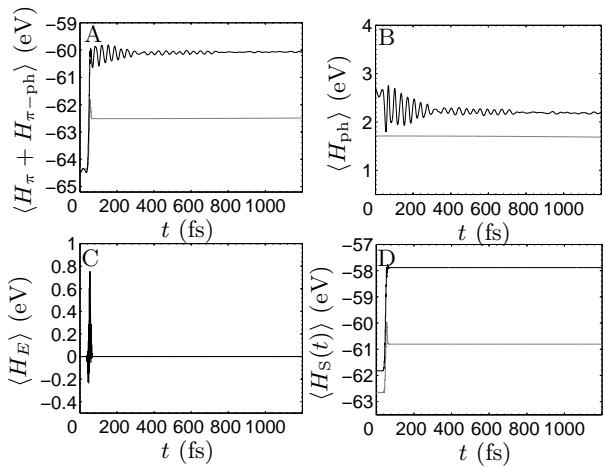}
\caption{Energy contributions during the dynamics of PA under the 
influence of pulse \textbf{f4} with $\hbar\w = 1.3$ eV and 
$\phi_{2\w}-2\phi_\w=0$ for the rigid (gray lines) and flexible 
(black lines) chain. Panel (A): the electronic kinetic energy 
(including electron-ion interactions), (B): purely nuclear 
energy, and (C): radiation-matter interaction. The total energy 
is shown in (D). }
     \label{fig:dynamics_ene_10fs}
\end{figure}
%----------------------------------------------------------------%

%----------------------------------------------------------------%
\begin{figure}[htbp]
\psfrag{A}[][]{A}
\psfrag{B}[][]{B}
\psfrag{t (fs)}[][]{$t$ (fs)}
\psfrag{ni}[][]{$\mean{n_\gamma(t)}$}
\psfrag{7th state}[][]{\footnotesize{$\,\gamma=7$}}
\psfrag{8th state}[][]{\footnotesize{$\,\gamma=8$}}
\psfrag{9th state}[][]{\footnotesize{$\,\gamma=9$}}
\psfrag{10th state}[][]{\footnotesize{$\,\gamma=10$}}
\psfrag{11th state}[][]{\footnotesize{$\,\gamma=11$}}
\psfrag{12th state}[][]{\footnotesize{$\,\gamma=12$}}
\psfrag{13th state}[][]{\footnotesize{$\,\gamma=13$}}
\psfrag{14th state}[][]{\footnotesize{$\,\gamma=14$}}
\centering
\includegraphics[width=0.5\textwidth]{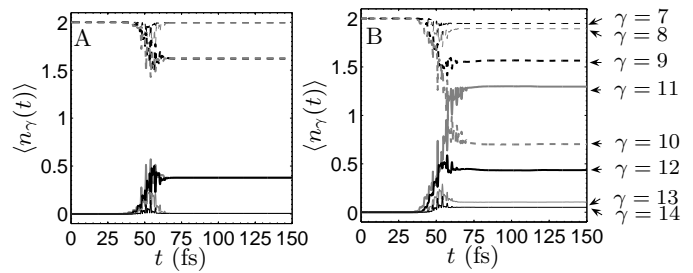}
\caption{Occupation numbers of the instantaneous eigenstates near the band
gap for the (A) rigid  and (B) flexible  chain during the dynamics induced by
pulse \textbf{f4} with $\hbar\w = 1.3$ eV and $\phi_{2\w}-2\phi_\w=0$. }
     \label{fig:pop_10fs}
\end{figure}
%----------------------------------------------------------------%

%----------------------------------------------------------------%
\begin{figure*}[htbp]
\psfrag{site (n)}[][]{site ($n$)}
\psfrag{energy (eV)}[][]{$\mean{\epsilon_\gamma}(t)$ (eV)}
\psfrag{t (fs)}[][]{$t$ (fs)}
\psfrag{8th }[][]{$\quad\gamma=8$}
\psfrag{9th}[][]{$\quad\gamma=9$}
\psfrag{10th}[][]{$\quad\gamma=10$}
\psfrag{11th}[][]{$\quad\gamma=11$}
\psfrag{12th}[][]{$\quad\gamma=12$}
\psfrag{13th state}{$\gamma=13$}
\centering
\includegraphics[width=0.8\textwidth]{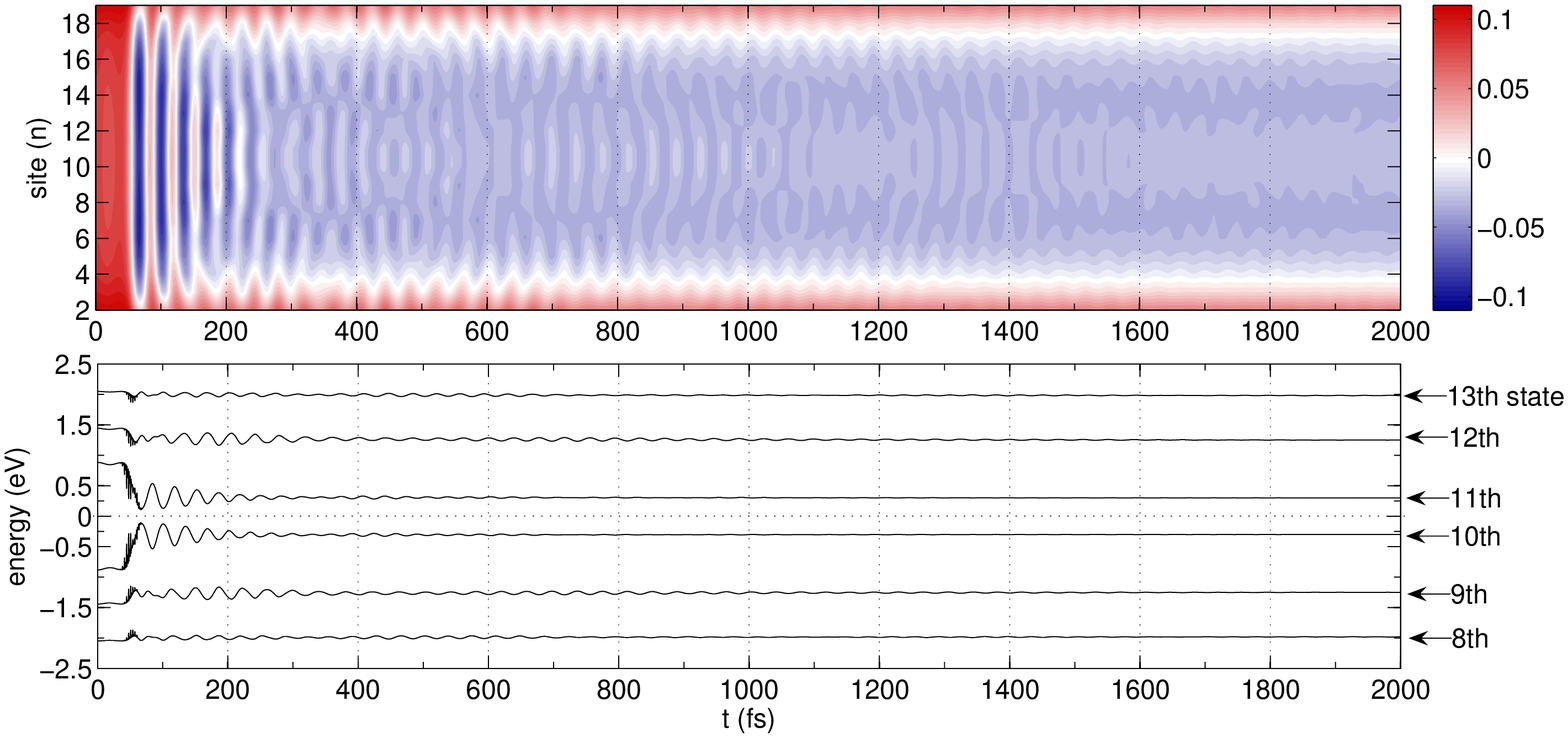}
\caption{Vibronic dynamics of PA induced by pulse \textbf{f4} with
$\hbar\w = 1.3$ eV and $\phi_{2\w}-2\phi_\w=0$. The upper panel 
shows contours of the bond length alternation parameter [defined 
in \eq{PAeq:bla}]. The lower panel, the mean instantaneous  
energies for the single-particle states near the energy gap.}
     \label{fig:bla_eigen}
\end{figure*}
%----------------------------------------------------------------%

%----------------------------------------------------------------%
\begin{figure}[htbp]
\psfrag{t (fs)}[][]{$t$ (fs)}
\psfrag{purity}[][]{$\text{Tr}\rho^2(t)$/$\text{Tr}\rho_\text{pure}^2$}
\psfrag{label}[][]{label $i$}
\centering
\includegraphics[width=0.4\textwidth]{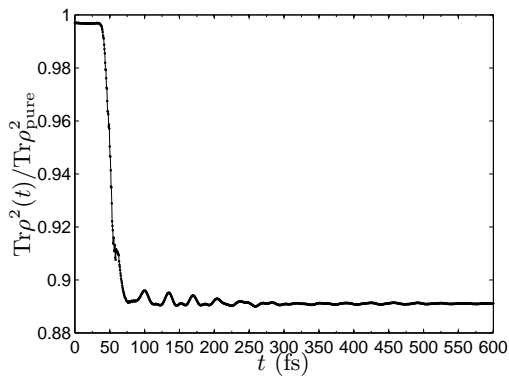}
\caption{Decoherence during the dynamics of the flexible chain induced by
pulse \textbf{f4} (10 fs wide, centered at 50 fs) with $\hbar\w=1.3$ eV and  $\phi_{2\w}-2\phi_\w=0$. }
     \label{fig:decoherence}
\end{figure}
%----------------------------------------------------------------%

The different contributions to the rigid and flexible chain 
energy  during and after photoexcitation with a 10 fs $\w+2\w$ 
pulse (laser field \textbf{f4} in 
Table~\ref{tbl:laserparameters}) with $\hbar\w=1.3$ eV are shown 
in \fig{fig:dynamics_ene_10fs}. Initially, the flexible chain has 
0.83 eV more energy than its rigid counterpart due to zero-point 
motion in the ground electronic state. Secondly, after 
photoexcitation, the rigid chain gains 1.85 eV from the laser 
while the average energy absorbed by the ensemble of flexible 
chains is 3.9 eV. This enhanced absorption of energy of the 
flexible chain is due to spectral broadening of states which form 
bands, with the broadening being especially large near the band 
gap  (recall Table~\ref{tbl:bandwidths}). As exemplified in 
\fig{fig:pop_10fs}, this results in a wider set of electronic 
levels being populated during photoexcitation.  In both cases, 
the occupation numbers and the single particle spectrum display 
the electron-hole symmetry of the SSH Hamiltonian.

Figure~\ref{fig:bla_eigen} shows the evolution of the bond length alternation
parameter  (upper panel) and the mean instantaneous orbital energies for
states near the energy gap (lower panel) of the flexible chain. The photoexcitation
of the chain initiates a complex highly nonlinear vibronic evolution.  The
lattice is set to vibrate and  large scale oscillations in the bond length
alternation are observed. By $t=67$ fs the chain has reversed the sense of its
dimerization  and establishes a coherent ``breathing'' oscillation of the bond
length alternation in which $\mean{r_n(t)}$ oscillates between the two senses
of dimerization. Several carbon-carbon stretching vibrational modes with
periods  between $32-40$ fs enter into the initial lattice dynamics.
The breathing motion, however, is not
long lived and the amplitude of the oscillations display an initial fast
exponential decay with a characteristic time-scale of $\sim 90-130$ fs,
followed by a slower damping. By $t\sim 400$ fs most of the vibrational modes
initially excited  have already decayed, and gradually the system approaches a
state of internal equilibrium through internal vibrational energy
redistribution.

These changes in the lattice geometry are reflected as changes in the
single-particle spectrum.   Upon initial photoexcitation the two states near
the energy gap  come very close to each other in less than 20 fs and the
band-gap energy fluctuates  following the breathing oscillation of the
bond-length alternation. A dip (crest) in the bond-length alternation
corresponds to a minimum (maximum) in the gap energy. Concurrent with
the dimerization pattern, the energy gap exhibits strong initial oscillations
at time scales between 30-40 fs  that rapidly decay,
followed by a slower damping. The band-gap energy  eventually reaches an
equilibrium value of $\sim 0.6$ eV.  The remaining energy levels also
oscillate on time-scales determined by the different vibrational modes of the
chain. We observe that states around the band-gap couple to vibrational modes
whose period is $\sim$38 fs, while states near the band edges couple to lower
frequencies modes.

An estimate of the electronic decoherence that occurs during and 
after the pulse is provided by  measuring the purity~\cite{xupei} 
$\text{Tr}{\rho^2}= \sum_{n,m} 
\overline{\rho}_{n,m}\overline{\rho}_{m,n}$, where 
$\overline{\rho}_{n,m}$ are the matrix elements of  the 
ensemble-averaged electronic reduced density matrix 
[\eq{PAeq:rdm}]. For pure systems with a $\rho_{n,m}(t)$ of the 
form in \eq{PAeq:rdmsimple}, as is the case of the rigid chain,  
$\text{Tr}\rho_{\text{pure}}^2$ is a constant equal to  $2\mc{N}$ 
where $\mc{N}$ is the number of electrons.  For mixed states 
$\text{Tr}\rho_{\text{mixed}}^2 \le 
\text{Tr}\rho_{\text{pure}}^2$ and its decay offers insight into 
the decoherence time scales introduced by the coupling to the 
vibrational modes.  Figure~\ref{fig:decoherence} shows the 
evolution of  $\text{Tr}\rho^2$ for the flexible chain.  The 
system begins in a state that is almost pure with 
$\text{Tr}{\rho^2}/\text{Tr}{\rho_{\text{pure}}^2}=0.997$.  Upon 
photoexcitation the purity displays a sharp drop at $\sim 40$ fs, 
followed by some modulation in time scales commensurate with the 
vibrational periodicities of the chain. These modulations decay 
rapidly and the purity stabilizes at 0.89 after $\sim 400$ fs. 

The dynamical behavior exemplified by our simulations describes 
quite well the main trends observed in recent sub-10 fs 
experiments that have been performed on PA~\cite{kobayashi} and 
PPV~\cite{lanzani} samples: initial fast decay of the breathing 
motion (order of 200 fs) followed by a slower damping. Previous 
studies of the excited state dynamics in PA did not capture the 
decay since they only consider the evolution of a single 
trajectory. However, the time scales that we observe for the 
energy gap fluctuations (32-38 fs) are somewhat larger than the 
ones reported for PA of  31 fs and 23 fs. This is a consequence 
of the parameters employed for the Hamiltonian.

We have observed that the basic features of the vibronic dynamics 
described above also apply to longer oligomers. However, in the 
long chain limit the two states near the energy gap can become 
degenerate, forming soliton-antisoliton pairs~\cite{SSH}, unlike  
to the case described above.

%----------------------------------------------------------------%
\subsubsection{The dynamics induced by a 300 fs pulse}
\label{PAsec:300fs}

%----------------------------------------------------------------%
\begin{figure}[htbp]
\psfrag{A}[][]{A}
\psfrag{B}[][]{B}
\psfrag{C}[][]{C}
\psfrag{D}[][]{D}
\psfrag{t (fs)}[][]{$t$ (fs)}
\psfrag{EE}[][]{$\mean{ H_\pi + H_{\pi-\text{ph}}}$ (eV)}
\psfrag{NE}[][]{$\mean{ H_\text{ph}}$ (eV)}
\psfrag{RM}[][]{$\mean{ H_E}$ (eV)}
\psfrag{TE}[][]{$\mean{ H_\ti{S}(t)}$ (eV)}
\centering
\includegraphics[width=0.5\textwidth]{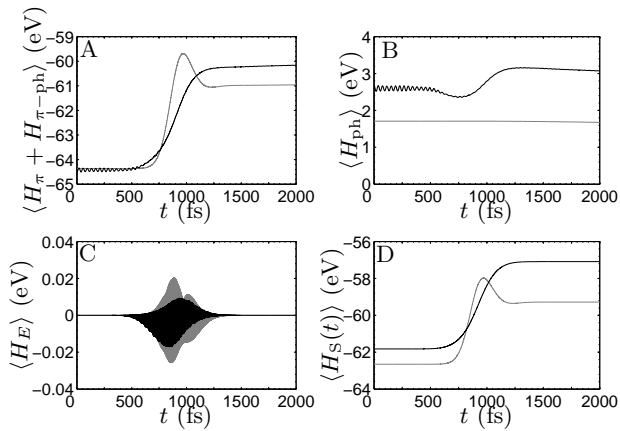}
\caption{Energy contributions during the dynamics of PA under the 
influence of pulse \textbf{f1} with $\hbar\w = 1.18$ eV and 
$\phi_{2\w}-2\phi_\w=0$. Panel (A)  electronic kinetic energy 
(including electron-ion interactions), (B) the purely nuclear 
energy, and (C) the radiation-matter interaction. The total 
energy is shown in (D). The gray and black  lines correspond to 
the rigid and flexible chains, respectively.}
     \label{fig:dynamics_ene_300fs}
\end{figure}
%----------------------------------------------------------------%

%----------------------------------------------------------------%
\begin{figure}[htbp]
\psfrag{A}[][]{A}
\psfrag{B}[][]{B}
\psfrag{time }[][]{$t$ (fs)}
\psfrag{ni}[][]{$\mean{n_\gamma(t)}$}
\psfrag{7th state}[][]{\tiny{$\,\gamma=7$}}
\psfrag{8th state}[][]{\tiny{$\,\gamma=8$}}
\psfrag{9th state}[][]{\tiny{$\,\gamma=9$}}
\psfrag{10th state}[][]{\tiny{$\,\gamma=10$}}
\psfrag{11th state}[][]{\tiny{$\,\gamma=11$}}
\psfrag{12th state}[][]{\tiny{$\,\gamma=12$}}
\psfrag{13th state}[][]{\tiny{$\,\gamma=13$}}
\psfrag{14th state}[][]{\tiny{$\,\gamma=14$}}
\centering
\includegraphics[width=0.5\textwidth]{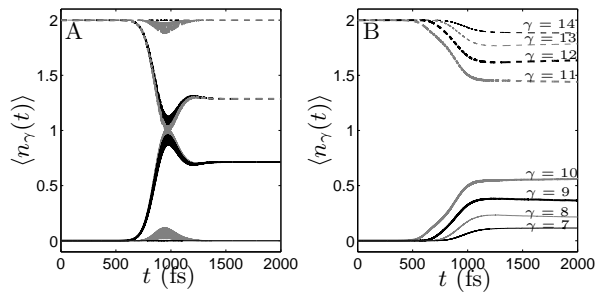}
\caption{Occupation numbers of the instantaneous eigenstates near the
energy gap during the dynamics induced by pulse \textbf{f1} with $\hbar\w = 1.3$ eV and
$\phi_{2\w}-2\phi_\w=0$. (A) The rigid chain; (B)  flexible case.}
     \label{fig:pop_300fs}
\end{figure}
%----------------------------------------------------------------%

%----------------------------------------------------------------%
\begin{figure*}[htbp]
\psfrag{site}[][]{site ($n$)}
\psfrag{energy}[][]{$\mean{\epsilon_\gamma}(t)$ (eV)}
\psfrag{time}[][]{$t$ (fs)}
\psfrag{7th state}[][]{\footnotesize{$\,\gamma=7$}}
\psfrag{8th state}[][]{\footnotesize{$\,\gamma=8$}}
\psfrag{9th state}[][]{\footnotesize{$\,\gamma=9$}}
\psfrag{10th state}[][]{\footnotesize{$\,\gamma=10$}}
\psfrag{11th state}[][]{\footnotesize{$\,\gamma=11$}}
\psfrag{12th state}[][]{\footnotesize{$\,\gamma=12$}}
\psfrag{13th state}[][]{\footnotesize{$\,\gamma=13$}}
\psfrag{14th state}[][]{\footnotesize{$\,\gamma=14$}}
\centering
\includegraphics[width=0.8\textwidth]{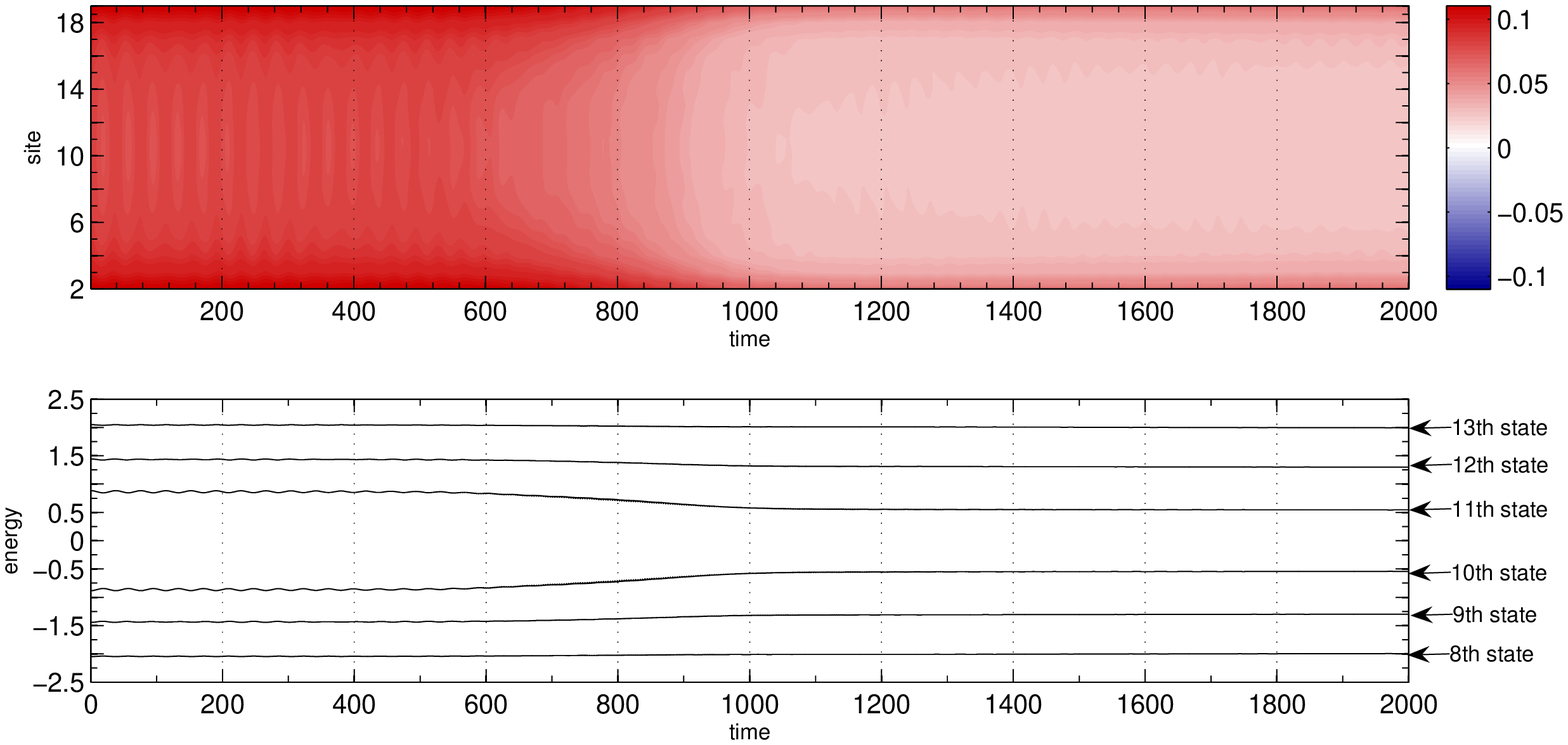}
\caption{The vibronic dynamics of PA induced by pulse \textbf{f1} with
$\hbar\w = 1.18$ eV and $\phi_{2\w}-2\phi_\w=0$. The upper panel shows
contours  of the bond length alternation parameter (defined in \eq{PAeq:bla}).
The lower panel shows the mean instantaneous  energies for the
single-particle states near the energy gap.}
     \label{fig:bla_eigen_300fs}
\end{figure*}
%----------------------------------------------------------------%

%----------------------------------------------------------------%
\begin{figure}[htbp]
\psfrag{t (fs)}[][]{$t$ (fs)}
\psfrag{purity}[][]{$\text{Tr}\rho^2(t)$/$\text{Tr}\rho_\text{pure}^2$}
\psfrag{label}[][]{label $i$}
\centering
\includegraphics[width=0.4\textwidth]{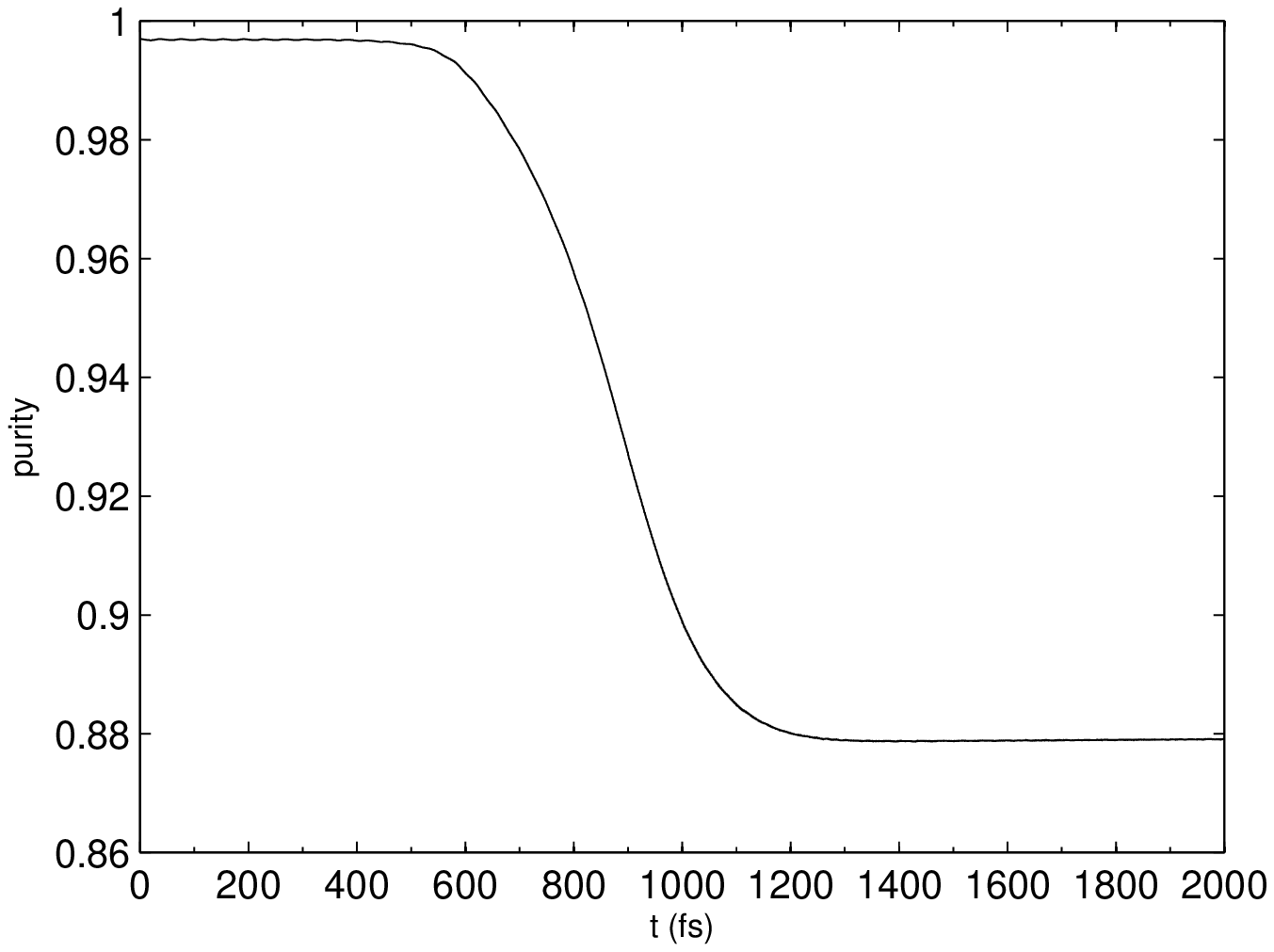}
\caption{Decoherence during the dynamics of the flexible chain induced by
pulse \textbf{f1} (300 fs wide, centered at 900 fs) with $\hbar\w=1.18$ eV and  $\phi_{2\w}-2\phi_\w=0$. }
     \label{fig:decoherence_300fs}
\end{figure}
%----------------------------------------------------------------%

Consider now dynamics under the longer (300 fs) pulse \textbf{f1} 
with $\hbar\w=1.18$ eV. Here the pulse time width  is  long 
compared with the typical optical-phonon vibrational modes. 
Figure~\ref{fig:dynamics_ene_300fs} and~\ref{fig:pop_300fs} show 
the different energy contributions to the chain during the 
dynamics and the instantaneous occupation numbers for states near 
the energy gap, respectively. As before, due to  broad  nature of 
its transition frequencies the  flexible chain absorbs 1.36 eV 
more energy from the field, with a wider range of electronic 
levels being populated during the process. The large-scale 
oscillations observed previously in the average electronic and 
nuclear energy for the flexible chain 
(\fig{fig:dynamics_ene_10fs}A-B) are no longer present.

The origin of this difference in  behavior is readily identified 
by considering  the bond length alternation and the single 
particle spectrum during the dynamics 
(Figure~\ref{fig:bla_eigen_300fs}). During the first 400 fs the 
bond-length alternation and the transition frequencies display 
minor oscillations. These oscillations arise because our initial 
state is not an exact eigenstate of the Hamiltonian due to ground 
state residual lattice anharmonicities. Upon photoexcitation the 
bond lengths in the chain slowly equalize in order to improve 
electronic delocalization along the chain that permits  
accommodating the  electrons that have been photoexcited.  After 
the pulse the  bond length alternation settles at $\sim0.03$\AA.  
These changes in the lattice geometry are accompanied by a 0.7 eV 
red shift of the energy gap. However, contrary to the dynamics 
under the 10 fs pulse,  the lattice does not establish a 
breathing motion. Instead the mean geometrical observables and 
spectroscopic quantities display a slow and steady change from 
the perfectly dimerized lattice to a chain with increased 
electronic delocalization. In essence,  the frequency width of 
the pulse is not broad enough to bring about a coherent breathing 
motion of the chain. The latter  requires participation of many 
vibrational states.  These features of the vibronic dynamics 
induced by 300 fs pulses also persist when considering longer 
oligomers.

Decoherence  occurs slowly and steadily during the time that the 
system is being driven by the laser field 
(\fig{fig:decoherence_300fs}). No modulations of  
$\text{Tr}\rho^2$ or further decoherence are observed after the 
pulse, and $\text{Tr}\rho^2/\text{Tr}\rho^2_\text{pure}$ 
stabilizes at $\sim 0.88$. Computations discussed below show that 
the electronic dephasing time is of the order of 2.5 fs. Hence, 
the limiting step in the decoherence process is not the rate at 
which lattice-induced decoherence occurs, but the rate at which 
the field is able to establish new superposition states that then 
rapidly  decohere.

\subsubsection{How fast is the electronic dephasing?}
\label{PAsec:decoherence}

%------------------------------------------------------------------%
\begin{figure}[htbp]
\centering
\psfrag{t (fs)}[][cb]{$t$ (fs)}
\psfrag{x/x0}[][]{$\frac{\mean{\mu(t)}}{\mean{\mu(0)}}$}
\psfrag{N=4}[][]{$N=4$}
\psfrag{N=20}[][]{$N=20$}
\psfrag{N=50}[][]{$N=50$}
\psfrag{N=100}[][]{$N=100$}
\includegraphics[width=0.50\textwidth]{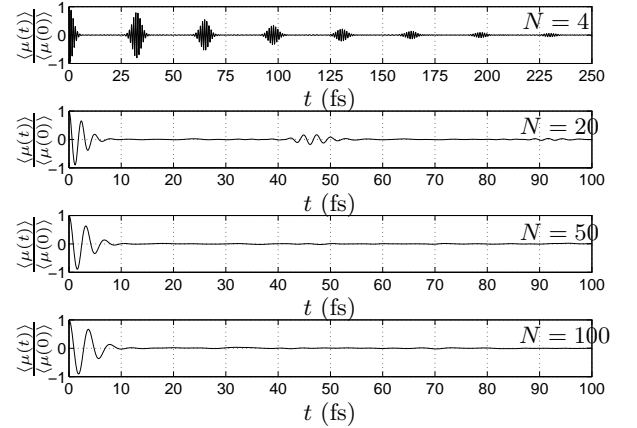}
\caption{Time dependence of the polarization $\mean{\mu(t)}$ for 
neutral flexible PA chains with $N$ sites. The initial state is 
a  superposition,  with equal coefficients, between the ground 
and first excited electronic state. The nuclei are taken to be 
initially in the ground-state configuration.  }
     \label{fig:coherence}
\end{figure}
%------------------------------------------------------------------%

A  measure of the timescale for the electronic dephasing induced 
by intramolecular vibrational motions can be obtained by 
following  the field-free evolution  of neutral chains  initially 
in an electronic superposition state. In a full quantum 
mechanical analysis, nuclear evolution on alternative electronic 
potential energy surfaces leads, in general, to electronic 
coherence loss. In our quantum-classical picture  the electronic 
dephasing is captured by the  distribution and evolution of  
orbital energies contained within the initial ensemble (recall 
Table \ref{tbl:bandwidths}).

A detailed calculation reveals the features of this dephasing.
Figure~\ref{fig:coherence} shows the time-dependence of the 
polarization for PA oligomers  initially prepared in  a 
superposition $\ket{\Psi(0)} = \frac{1}{\sqrt{2}}( \ket{\ti{G}} + 
\ket{\ti{E}})$, where  $\ket{\ti{G}}$ and $\ket{\ti{E}}$ are the 
$N$-particle ground and first-excited electronic states. The 
initial nuclear distribution is taken to be the one obtained in 
the ground electronic surface, and  $\ket{\ti{E}}=  
c_{\epsilon''+1,s'}^\dagger c_{\epsilon'',s'}\ket{\ti{G}}$  is 
generated  by  instantaneously promoting an electron from the 
HOMO  to the LUMO orbital of the chain, so that 
$\epsilon''=N/2$.  For chains with  4 sites  the polarization  
displays a fast initial decay, and observes recurrences every 
$\sim 30$ fs. In a full quantum mechanical analysis these 
recurrences arise from the time dependence of the overlap of the 
nuclear wavefunctions in the ground and excited electronic 
states. In this quantum-classical picture, the recurrences are 
captured by the time-dependence of the orbital energies during 
the dynamics. Note that between consecutive recurrences the 
amplitude of the polarization gets reduced  and $\mean{\mu(t)}$  
dies in hundreds of femtoseconds. For longer chains these 
recurrences are less important  and the dephasing time is 
basically determined by the initial decay of the polarization 
which occurs in less than  10 fs, with a characteristic timescale of 2.5 fs.
These time-scales are in agreement with  
electronic dephasing timescales that have been determined in 
other systems using a fully quantum treatment~\cite{rossky}.

%----------------------------------------------------------------%
\subsection{Laser-induced symmetry breaking}
\label{PAsec:control}

%----------------------------------------------------------------%
\begin{figure*}[htbp]
\psfrag{A}[][]{A}
\psfrag{B}[][]{B}
\psfrag{C}[][]{C}
\psfrag{A'}[][]{A'}
\psfrag{B'}[][]{B'}
\psfrag{C'}[][]{C'}
\psfrag{X(t)}[][]{$\mean{\mu(t)}/|e|$ (\AA)}
\psfrag{t (fs)}[][]{$t$ (fs)}
\psfrag{C(t)}[][]{$\mean{C(t)}/|e|$}
\centering
\includegraphics[width=0.7\textwidth]{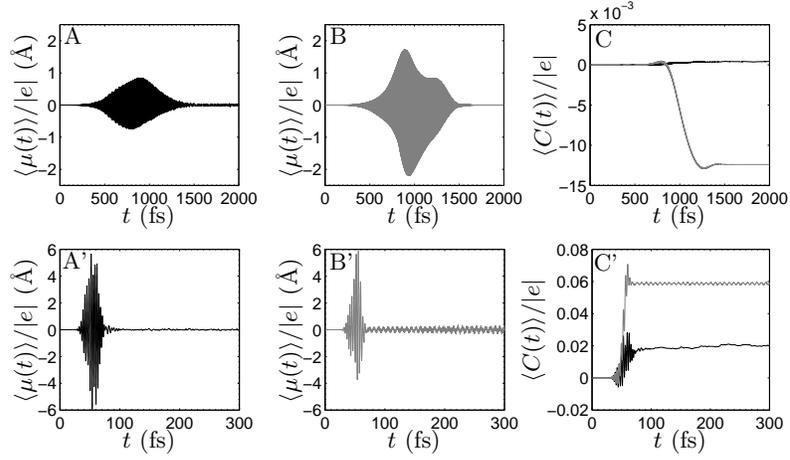}
\caption{First moment of the charge distribution   during and 
after photoexcitation with a symmetry breaking $\omega + 2\omega$ 
laser field. (A) the flexible and (B) the rigid chain under the 
influence of  pulse \textbf{f1}  with $\hbar\w =1.18$ eV and 
$\phi_{2\w}-2\phi_\w=0$. Panel (C)  the resultant cumulative 
dipoles  $\mean{C(t)}$ (black line, flexible chain; gray line, 
rigid case). Parts (A')-(C')  as in  (A)-(C) but for the 
\textbf{f4} pulse with $\hbar\w = 1.3$ eV and 
$\phi_{2\w}-2\phi_\w=0$.   }
     \label{fig:pos_figs}
\end{figure*}
%----------------------------------------------------------------%
%----------------------------------------------------------------%
\begin{figure*}[htbp]
\psfrag{A}[][]{A}
\psfrag{B}[][]{B}
\psfrag{C}[][]{C}
\psfrag{D}[][]{D}
\psfrag{pulse f1}[][]{pulse \textbf{f1}}
\psfrag{pulse f2}[][]{pulse \textbf{f2}}
\psfrag{pulse f3}[][]{pulse \textbf{f3}}
\psfrag{pulse f4}[][]{pulse \textbf{f4}}
\psfrag{control}[][]{$\mean{C(\infty)}/|e|$}
\psfrag{hbaromega}[][]{$\hbar\w$ (eV)}
\centering
\includegraphics[height=0.7\textwidth, angle= -90]{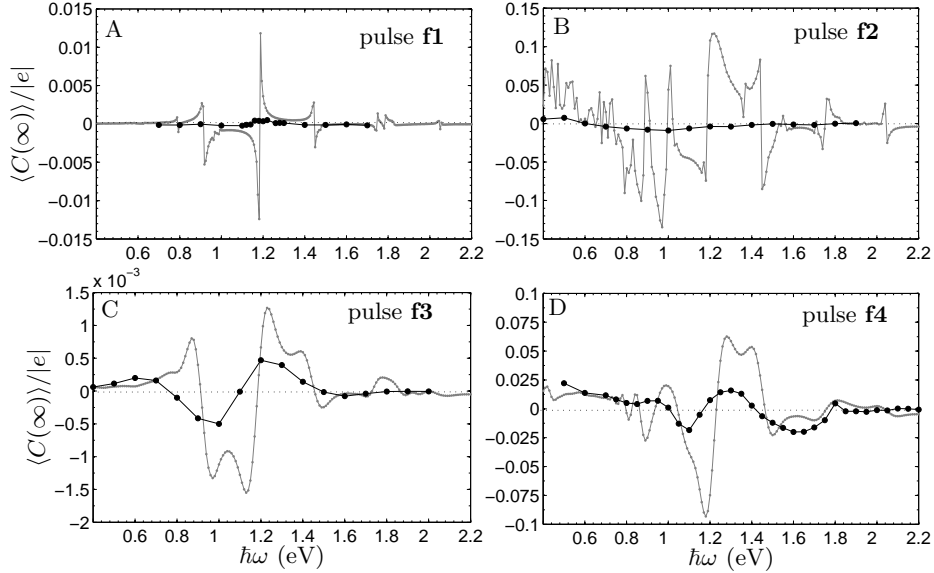} 
\caption{ Frequency dependence of 
the degree of symmetry breaking generated by an $\w+2\w$ field in 
rigid (gray) and flexible (black)  PA oligomers with 20 sites. 
The initial energy gap of the oligomer is $2\Delta=1.8$ eV. The 
plots show the asymptotic cumulative dipole $\mean{C(\infty)}$  
after photoexcitation with pulses (A) \textbf{f1}, (B) 
\textbf{f2}, (C) \textbf{f3} and (D) \textbf{f4}, using 
$\phi_{2\w}-2\phi_\w=0$. }
     \label{fig:all_freq_dependence}
\end{figure*}
%----------------------------------------------------------------%
%----------------------------------------------------------------%
\begin{figure}[htbp]
\psfrag{A}[][]{\footnotesize{A}}
\psfrag{B}[][]{\footnotesize{B}}
\psfrag{C}[][]{\footnotesize{C}}
\psfrag{control}[][]{\footnotesize{$\mean{C(\infty)}/|e|$}}
\psfrag{phase}[][]{\footnotesize{$(\phi_{2\w} - 2\phi_\w)/\pi$}}
\centering
\includegraphics[width=0.5\textwidth]{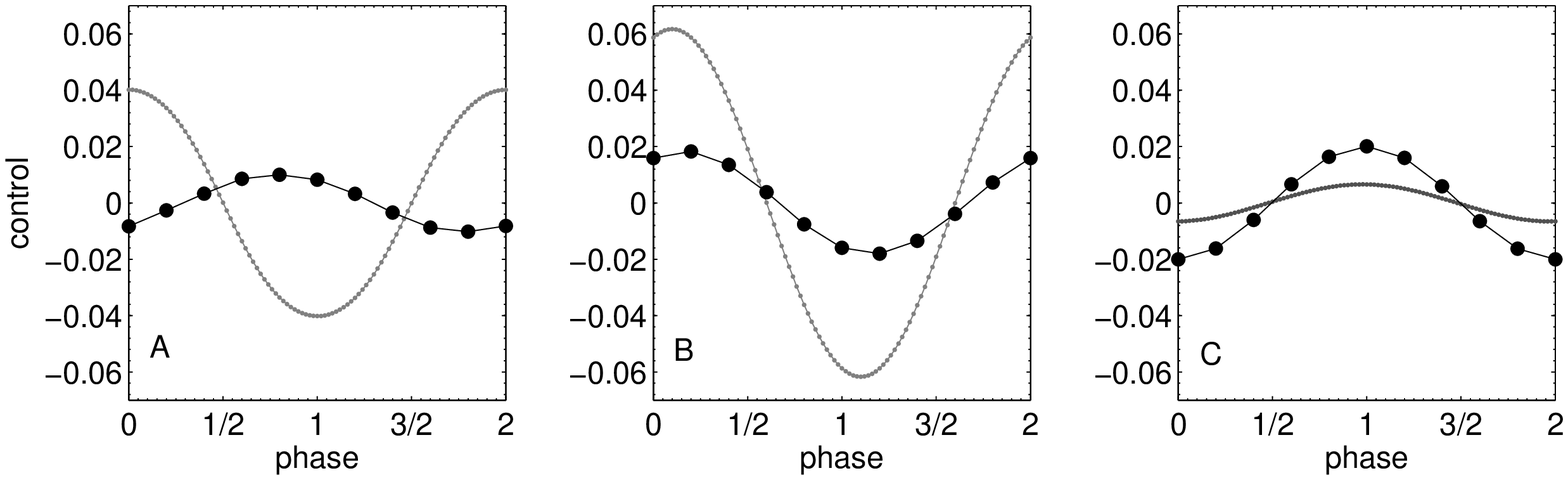}
\caption{The degree of symmetry breaking induced by an $\w+2\w$ field in
rigid (gray) and flexible (points) PA, as a function of the relative phase
$\phi_{2\w} -2\phi_\w$ of the laser. The figures show the mean cumulative
dipole $\mean{C(\infty)}$ after photoexcitation with: (A)  pulse
\textbf{f2} with $\hbar\w=0.9$ eV;  (B)  pulse \textbf{f4} with
$\hbar\w=1.3$ eV; and (C) pulse \textbf{f4} with $\hbar\w=1.6$ eV. }
\label{fig:phase_dependence}
\end{figure}
%----------------------------------------------------------------%

In the previous sections we identified  important ways in 
which the vibronic couplings influence the photoinduced dynamics 
of $\pi$-conjugated systems.  The coupling introduces: (i) 
broadening of the electronic transitions (see 
Table~\ref{tbl:bandwidths}, and Figs.~\ref{fig:pop_10fs} 
and~\ref{fig:pop_300fs}),  (ii) pronounced changes in the mean 
single-particle spectrum (see Figs.~\ref{fig:bla_eigen} 
and~\ref{fig:bla_eigen_300fs}), (iii) internal relaxation 
mechanisms, and (iv) ultrafast dephasing (see 
Figs.~\ref{fig:decoherence}, \ref{fig:decoherence_300fs} and 
\ref{fig:coherence}). Because of this, when using  lasers  to 
control a dynamical process in this class of systems, the laser 
frequencies may become detuned from the desired transition and 
the control may be diminished by broadening of the electronic 
levels and internal relaxations. Here we investigate the extent  
to which these additional complexities  affect the ability  to 
induce symmetry breaking using the $\w$ vs. $2\w$ coherent 
control scenario.

 The results presented in this section focus on  
effects that depend on the relative phase  of the laser field. 
This is because symmetry breaking contributions that depend on 
the carrier envelope phase are typically difficult to control 
(although not impossible, see e.g., Ref.~\onlinecite{kling}) 
since it requires an experimental arrangement that permits 
locking the carrier envelope phase of the incident radiation and 
that controls the position of the center of mass of the molecule 
with respect to the laboratory reference frame. This contrasts 
with relative-phase control that is unaffected by the center of 
mass motion and  only requires manipulating the relative phase 
between the two central frequency components of the $\w+2\w$ 
field.

We consider the evolution of  rigid and flexible chains under the 
influence of the laser pulses  in 
Table~\ref{tbl:laserparameters}, for different laser frequencies. 
The parameters chosen for the laser pulses  encompass  four 
illustrative  cases: dynamics induced by weak and moderately 
strong pulses with time envelopes that are either  short or  long 
with respect to the typical 30-40 fs optical-phonon vibrational 
period.  Specifically, pulses \textbf{f1} and \textbf{f2} induce 
weak and strong-field dynamics, respectively, when the time-width 
of the pulse is relatively long (300 fs). By contrast, pulses 
\textbf{f3}  and \textbf{f4} offer insight into the weak- and 
strong-field control when the laser pulses are short  (10 fs), 
and of the order of the electronic dephasing time, permitting 
only a limited degree of excited state dynamics during the pulse.

Consider first the dynamics in the weak-field/long-pulse 
(\textbf{f\,1}) and strong-field/short-pulse (\textbf{f\,4}) 
regimes. The relevant observable here is the polarization of the 
chain  $\langle\mu(t)\rangle$ [\eq{PAeq:x}] which  quantifies the 
symmetry breaking. Figure~\ref{fig:pos_figs} shows representative 
results. Initially $\mean{\mu}=0,$ reflecting the symmetric 
distribution of electrons over the chain. As soon as the laser is 
turned on, dipoles are induced in the system, oscillating with 
the frequencies of the field and various harmonics. The harmonic 
of interest  is the zero-frequency (DC) component, appearing at 
the 3rd, 5th, etc.  order response to the $\w+2\w$ field. These 
harmonics can be extracted by integrating the signal over time 
[see \eq{PAeq:cumx}], yielding the cumulative dipoles  
$\mean{C(t)}$ shown in \fig{fig:pos_figs}C and \fig{fig:pos_figs} 
C\;$^\prime$. We denote the mean asymptotic value of 
$\mean{C(t)}$ after the pulse  as $\mean{C(\infty)}$.  As can be 
seen, all symmetry breaking effects induced by the  $\w+2\w$ 
field are achieved  while the system is being driven by the laser 
field.   Once the field is turned off, the zero-frequency 
component disappears so that $\mean{C(t)}$ becomes constant and 
net dipoles no longer persist. This is because systems with 
discrete, non-degenerate, spectra cannot sustain net dipoles 
after the pulse.

In the rigid case the zero-frequency term constitutes an 
important component of $\mean{\mu(t)}$ and both \textbf{f\,1} and 
\textbf{f\,4}  pulses generate a net dipole while the pulse is 
on.  By contrast,   when pulse \textbf{f\,1} is applied on the 
flexible chain (\fig{fig:pos_figs} upper panels) the 
electron-vibrational couplings  mute most of the effect. However, 
by applying a shorter pulse (lower panels) it is possible to 
generate net dipoles that are of the same order of magnitude as 
the ones observed for the rigid chain.  In essence, by using   
shorter pulses one is limiting the detrimental effects that the 
excited state vibronic couplings exert on the control.

A complementary perspective on the influence  of the vibronic 
couplings on the photoinduced control is obtained by  considering 
the dependence of the symmetry breaking effect on the laser 
frequency $\w$. Figure~\ref{fig:all_freq_dependence} shows the 
asymptotic cumulative dipoles  $\mean{C(\infty)}$ observed in 
flexible and rigid chains after photoexcitation with pulses 
\textbf{f1}-\textbf{f4} for different laser frequencies. In the 
rigid case, the $\w$ vs. $2\w$ coherent control scenario is very 
robust, inducing  net dipoles at most driving frequencies and 
with pulses of any time-width and intensity. The effect shows 
sharp resonances at selected frequencies.  The origin of these 
resonances can be identified by comparing the control map to the 
single particle spectrum of the rigid chain 
(Table~\ref{tbl:bandwidths}). Consider,  for instance, the 
resonances displayed in \fig{fig:all_freq_dependence}A at 
$\hbar\w=$0.9, 1.18 and 1.42 eV. At these frequencies the 
quantity $2\w$ coincides with the HOMO$\to$LUMO, HOMO$\to$ LUMO+1 
(or HOMO-1$\to$ LUMO), and HOMO-1$\to$ LUMO+1 transition 
frequencies, respectively.  By applying stronger pulses (e.g. 
\fig{fig:all_freq_dependence}B) it is possible to exploit high 
order multiphoton processes  to generate dc terms, resulting in a 
complicated frequency dependence of the degree of control.

The control map in the flexible chain is remarkably different. In 
flexible chains, in order to generate appreciable symmetry 
breaking it is necessary to work with either sufficiently short 
pulses, so that only limited excited state dynamics occurs during 
the pulse (Figs.~\ref{fig:all_freq_dependence}C-D), or to apply 
stronger fields, as in \fig{fig:all_freq_dependence}B, in which 
the evolution imposed by the field becomes dominant. When using 
short pulses, the resonance structure previously observed in the 
rigid-chain control case is partially maintained.  However, in 
the case of  long pulses  the fine features observed in the rigid 
chain are washed away by the broadening of the electronic 
transitions, with only the rough features in the control map 
being maintained. Interestingly, when applying short pulses at 
selected frequencies it is possible to obtain a higher degree of 
symmetry breaking in the flexible chain than in the rigid chain.  
However, the maximum control attainable (as a function of $\w$) 
is always attenuated by the  electron-vibrational couplings.

Importantly, in all cases the sign and magnitude of symmetry 
breaking can be manipulated  by varying the relative phase 
between the two components of the field. 
Figure~\ref{fig:phase_dependence} exemplifies this dependence for 
selected frequencies and pulses. The electron-vibrational 
couplings in the system do not destroy the phase control of the 
electronic dynamics, but merely attenuate its magnitude. 
It follows then that it is possible to use the techniques 
of coherent control to manipulate electronic dynamics even in the 
presence of significant vibronic couplings. The attenuation of the 
effect can be overcome by using stronger pulses or by working 
with laser pulses that are short or comparable with the 
time-scale for electronic dephasing. Similar results were found 
in sample studies on longer chains.

\section{Conclusions}
\label{PAsec:conclusions}

We have investigated the possibility of inducing  electronic 
transport in conjugated polymers by irradiating the sample with 
$\w+2\w$ lasers. To do so, we have followed the highly nonlinear 
dynamics of the electronic and vibrational degrees of freedom in 
\emph{trans}-polyacetylene oligomers in the presence of  $\w+2\w$ 
laser pulses of different durations and intensities, and for a 
variety of frequencies. The simulations are performed in the mean 
field approximation, in which the nuclei are treated classically 
and the electrons  quantum mechanically. The dynamics is followed 
for an ensemble of lattice initial conditions obtained by 
sampling the ground-state nuclear Wigner distribution function in 
the harmonic approximation. 

Several important ways in which the electron-vibrational 
couplings in $\pi$-conjugated systems modify the photoinduced 
dynamics have been identified. The vibronic couplings introduce 
broadening of the electronic transitions and  can cause ultrafast 
dephasing in the electronic dynamics, as well as pronounced 
changes in the mean single-particle spectrum and IVR. 

The simulations reveal that the vibronic couplings can have 
strong detrimental effects on the control. In fact, it is 
difficult to induce and control electronic transport when the 
laser pulses are longer than the typical carbon-carbon 
vibrational period of $\sim 30$ fs, unless the pulse intensities 
are strong enough to completely dominate  the dynamics. However, 
even in the presence of significant vibronic couplings, limited laser 
control of the electronic dynamics is possible via the use of 
pulses whose durations are comparable to the electronic dephasing 
timescale ($\sim 10$ fs).

These results  provide insight into the way coherent control works
in large molecular systems with significant  electron-vibrational
couplings, and offer a characterization of the time-scale of 
decay of photoinduced breathers in conjugated polymers. We expect 
that the set of phenomena investigated here will be of generic 
importance in a wide range of materials characterized by significant 
electron-vibrational couplings.

%----------------------------------------------------------------%
\bibliography{608825JCP}

\end{document}